\documentclass
[12pt]{article}
\usepackage{latexsym}
\usepackage{amsfonts}
\usepackage{amsmath}
\usepackage{amssymb}

\begin{document}
\title{Two-dimensional Riemannian and Lorentzian geometries from second order ODEs}
\author{Emanuel Gallo \thanks{egallo@fis.uncor.edu}\\\\ \footnotesize{\textit{FaMAF, Universidad Nacional de
C\'{o}rdoba, 5000 C\'{o}rdoba, Argentina}}}

\date{} \maketitle

\begin{abstract}
In this note we give an alternative geometrical derivation of the
results recently presented by Garc\'{i}a-God\'{i}nez, Newman and
Silva-Ortigoza in \cite{GNS1} on the class of all two-dimensional
riemannian and lorentzian metrics from 2nd order ODEs which are in
duality with the two dimensional Hamilton-Jacobi equation. We show
that, as it happens in the Null Surfaces Formulation of General
Relativity, the W\"unschmann-like condition can be obtained as a
requirement of a vanishing torsion tensor. Furthermore, from these
second order ODEs we obtain the associated Cartan connections.

\end{abstract}

\newpage
\pagenumbering{arabic}
\section{Introduction}
In a couple of recent works, Garc\'{i}a-God\'{i}nez, Newman and
Silva-Ortigoza (GNS) presented the (pseudo)-riemannian geometries
which are hidden in certain class of differential equations (those
which satisfy a W\"unschmann-like condition, $I_{GNS}=0$). In the
first of these works \cite{GNS1}, they studied how to obtain all
two-dimensional riemannian and lorentzian metrics from certain
class of 2nd order ODEs. Furthermore, in \cite{GNS2}, they
extended their work and showed how to get all three-dimensional
metrics from certain class of three second-order PDEs and also
from a certain class of third order ODEs. From now on, we will say
that these equations are in the GNS-class.

The special status of these ODEs and PDEs is that they are in
duality with the Hamilton-Jacobi equation. For example, if we have
a 2nd order ODE in the GNS-class,
 \begin{equation}
u''=\Lambda(u,u',s), \end{equation} and if we know a solution
$u=Z(x^a,s)$, with $x^a=(x^1,x^2)$ integration constants, then
this solution automatically satisfies the two-dimensional
Hamil\-ton-Jacobi equation:
\begin{equation}
g^{ab}\nabla_aZ\nabla_bZ=1,\end{equation}
 where $\nabla_a$ means a differentiation with respect to $x^a$, and
$g^{ab}$, is a (pseudo)-riemannian metric constructed from
$\Lambda$ and its derivatives.

All these problems share  similar characteristics with the problem
of the Null Surface Formulation (NSF) of General Relativity in
three and four dimension (\cite{KN}-\cite{Forni}). In NSF, from
certain class of differential equations, known as the W\"unschmann
class, one can construct all three and four dimensional conformal
lorentzian metrics. The 3-dim conformal metrics are obtained from
a class of third order ODE,
\begin{equation}
u'''=F(u,u',u'',s),\label{nsf3d}
\end{equation}
with $F$ satisfying the so called W\"unschmann condition $I[F]=0$.
Likewise, the 4-dim metrics are obtained from a pair of second
order of PDEs,
\begin{eqnarray}
u_{ss}&=&\Lambda(s,s^*,u,u_s,u_{s^*},u_{ss^*}),\nonumber\\
u_{s^*s^*}&=&\Lambda^*(s,s^*,u,u_s,u_{s^*},u_{ss^*}).\label{nsf4d}
\end{eqnarray} where $s$ and $s^*$ are complex variables, and
$\Lambda$ and $\Lambda^*$ satisfy the generalized W\"unschmann
condition $W[\Lambda,\Lambda^*]=0$ and its complex conjugate. It
can be shown that $I[F]$, and $W[\Lambda,\Lambda^*]$, are
invariant under contact transformations~\cite{Ch,Nurtres,FKN}.
 Again, these equations are in duality with another equation, namely the eikonal equation,
\begin{equation}
g^{ab}\nabla_aZ\nabla_bZ=0.\end{equation} and the level surfaces
of the solution $u=Z(x^a,s)$ to (\ref{nsf3d}), or $u=Z(x^a,s,s^*)$
to (\ref{nsf4d}) are null surfaces of the respective metrics that
they generate.

  In NSF, the W\"unschmann condition can be obtained in several ways~\cite{KN,SCT,STN,GKNP}. Two of these were used by
GNS to obtain the W\"unschmann-like condition for differential
equations in duality with the Hamilton-Jacobi equation. There
exist a third method which is used in NSF, the torsion-free method
, and from which one can get not only the W\"unschmann class and
its respective metrics, but also more geometrical structures
associated to the equations, in particular all normal Cartan
conformal connections~\cite{FKNN,GKNP}. In this note we show that
this method can also be applied to the problem of
(pseudo)-riemannian metrics discussed by GNS. In particular, we
show how the torsion free condition restrict the class of second
order ODEs to those belong to the GNS-class and such that we get
all two-dimensional riemannian and lorentzian metrics, and
respective Cartan connections.

  In section 2 we briefly present the notation and basic concepts about the geometry of  2nd order ODEs. In section 3,
we show how to get the GNS-class from the torsion free condition,
and  construct the associated Cartan connections. Finally, in the
conclusions, we discuss the extension of this approach to the
problem of the GNS class of third order ODE's.

\section{Notation and basics notions}
Let the second order ODE be
\begin{equation}
u^{\prime \prime }=\Lambda(u,u^{\prime},s) \label{second-order1}
\end{equation}
where $s\in\mathbb{R}$ is the independent variable, and the primes
denote derivative of the dependent variable $u$ with
respect to $s$.\\
On the jet-space $J^1$ with local coordinates $(s,u,u^{\prime})$
we consider the Pfaffian system $\mathcal{P}$
\begin{eqnarray}
\omega^1&=&du-u'\; ds,\\
\omega^2&=&du'-\Lambda\; ds.\label{omega}
\end{eqnarray}
Local solutions of (\ref{second-order1}) are in one to one
correspondence with integral curves $\gamma:
\mathbb{R}\longrightarrow J^1$ of $\mathcal{P}$ satisfying
$\gamma^*\,ds\neq0$. These curves are generated by the vector
field on $J^1$ given by
\begin{equation}
e_s\equiv D=\frac{\partial}{\partial s}+u'\frac{\partial}{\partial
u}+ \Lambda\frac{\partial}{\partial u'}.
\end{equation}
We will restrict the domain of $\Lambda$ to a open neighborhood
$U$ of $J^1$ where $\Lambda$ is $C^\infty$ and the Cauchy problem
is well posed. Then, it follows from Frobenius theorem that the
solution space $M$ is a two-dimensional $C^\infty$ manifold, and
we will denote a given local coordinates system on it by
$x^a=(x^1,x^2)$. It means that we can construct a map
$Z:M\times\mathbb{R}\rightarrow\mathbb{R}$, $u=Z(x^a,s)$, such
that by a given $x^a_0\in M$ the map  $u=Z(x^a_0,s)$ is a solution of (\ref{second-order1}).\\
Then, if on $M\times\mathbb{R}$ we define a pfaffian system
$\mathcal{S}$ generated  by
\begin{eqnarray}
\beta^1&=&Z_a\,dx^a,\nonumber\\
\beta^2&=&Z'_a\,dx^a,\nonumber
\end{eqnarray}
(where primes means derivatives on $s$, and  $Z_a=\partial_a Z$),
it follows  that there exist a diffeomorphism $\zeta:
J^1\rightarrow M\times\mathbb{R}$ which pulls back the pfaffian
system $\mathcal{S}$ on the system $\mathcal{P}$, i.e.
\begin{equation}
\zeta^*\mathcal{S}=\mathcal{P}. \end{equation} We will make use of
this diffeomorphism later.

\section{Riemannian and lorentzian geometries from second order ODEs}
From $\omega^1$, $\omega^2$ which generate the pfaffian system
$\mathcal{P}$, we construct the following one-forms:
\begin{eqnarray}
\theta^1&=&\frac{1}{\sqrt{2}}\left (\omega^1+a\,\omega^2 \right ),\label{theta1}\\
\theta^2&=&\frac{1}{\sqrt{2}}\left (\omega^1-a\,\omega^2\right
),\label{theta2}
\end{eqnarray}
where $a=a(s,u,u')$ is a non-vanishing function to be determined.
Next, we construct a degenerate metric on $J^1$,
\begin{equation}
h(u,u',s)=2\theta^{(1}\otimes\theta^{2)}=\eta_{ij}\theta^i\otimes\theta^j,
\end{equation}
where
$$
\eta_{ij}=\left (\begin{array}{cc}
0&1\\
1&0\\
\end{array}
\right ).$$ \\
Note that if $a^2>0$, then $\theta^1$ and $\theta^2$ behave as
null real vectors, and if $a^2<0$, they are complex null
vectors.\\
Let be $\omega^i_j$ a connection such that :\\\\
  A. The connection is skew-symmetric
 \begin{equation}
 \omega_{ij}=\omega_{[ij]}, \label{antisimet}
 \end{equation} where
 $\omega_{ij}=\eta_{ik}\omega^k\,_j$.\\\\
  B. The one-forms $\theta^1$ and $\theta^2$ satisfy the Cartan's torsion-free first structure equations
 \begin{equation}
 T^i\equiv d\theta^i+\omega^i\,_j\wedge\theta^j=0.\label{free-torsion}
 \end{equation}\\
Now, we state and prove the following theorem:\\
\textbf{Theorem:}\textit{   The Torsion-free condition on the
skew-symmetric connection :\\
1. Uniquely determines the connection, with the only non vanishing
component given by
\begin{equation}
\omega_{[12]}=-\frac{1}{\sqrt{2}}(\ln
a)_u\theta^1+\frac{1}{\sqrt{2}}(\ln a)_u\theta^2+\frac{1}{a}ds.
\end{equation}
2. Uniquely determines the function $a$ in terms of $\Lambda$:
\begin{equation}
a^2=\frac{1}{\Lambda_u}.
\end{equation}\\
3. Impose a W\"unschmann-like condition on $\Lambda$:
\begin{equation}
I_{GNS}=Da+a\Lambda_{u'}=0.
\end{equation}}\\
\textbf{Proof:}  From (\ref{theta1}) and (\ref{theta2}) we have:
\begin{eqnarray}
d\theta^1=-\frac{1}{\sqrt{2}a}a_u
\theta^1\wedge\theta^2-\frac{1}{2a}\left
(1+Da+a^2\Lambda_u+a\Lambda_{u'}\right )\theta^1\wedge
ds\nonumber\\
-\frac{1}{2a}\left (-1-Da+a^2\Lambda_u-a\Lambda_{u'}\right
)\theta^2\wedge ds,\\
d\theta^2=\frac{1}{\sqrt{2}a}a_u
\theta^1\wedge\theta^2+\frac{1}{2a}\left
(-1+Da+a^2\Lambda_u+a\Lambda_{u'}\right )\theta^1\wedge
ds\nonumber\\
+\frac{1}{2a}\left (1-Da+a^2\Lambda_u-a\Lambda_{u'}\right
)\theta^2\wedge ds.
\end{eqnarray}
The condition of free-torsion (\ref{free-torsion}) reads
\begin{eqnarray}
d\theta^1-\omega_{[12]}\wedge\theta^1=0,\\
d\theta^2+\omega_{[12]}\wedge\theta^2=0,
\end{eqnarray}
and by solving  these equations we get:
\begin{eqnarray}
\omega_{[12]}&=&-\frac{1}{\sqrt{2}}(\ln
a)_u\theta^1+\frac{1}{\sqrt{2}}(\ln a)_u\theta^2+\frac{1}{2a}\left
(1+Da+a^2\Lambda_u+a\Lambda_{u'}\right)ds,\nonumber\\\label{conexion}
\end{eqnarray}
together to the three conditions:
\begin{eqnarray}
\left(-1-Da+a^2\Lambda_u-a\Lambda_{u'}\right)&=&0, \label{cond1}\\
\left(-1+Da+a^2\Lambda_u+a\Lambda_{u'}\right)&=&0, \label{cond2}\\
\left(Da+a\Lambda_{u'}\right)&=&0. \label{cond3}
\end{eqnarray}
Finally, from (\ref{conexion}), and the conditions (\ref{cond1}),
(\ref{cond2}), (\ref{cond3}) we get the results stated in the
theorem. Q.E.D.\\\\
Note now, that with the map $\zeta: J^1\rightarrow
M\times\mathbb{R}$ discussed in section 2, we have a 1-parameter
family of (pseudo)-riemannian metrics in the solution space $M$
(riemannian metrics if $\Lambda_u<0$, and lorentzian metrics if
$\Lambda_u>0$), i.e: we have the following family of metrics:
\begin{equation}
g(x^a,s)=(\zeta^{-1})^*\,h,
\end{equation}
or written in coordinates:
\begin{equation}
g(x^a,s)=\beta^1\otimes\beta^1-\frac{1}{\Lambda_u}\,\beta^2\otimes\beta^2=
\left[Z_aZ_b-\frac{1}{\Lambda_u}\,Z'_aZ'_b\right]dx^adx^b.
\end{equation}
in fact, all they are equivalents, because it is easy to show that
$h$ satisfy:
\begin{equation}
\pounds_{e_s}h=0.
\end{equation}
Finally, let us collect the one-forms $\theta^i$ and $\omega^i_j$
into the matrix-valued one-form
$$\omega_c=\left(
\begin{array}{ccc}
  0 & 0 & 0 \\
  \theta^1 & -\omega_{[12]}&0 \\
  \theta^2 & 0 &\omega_{[12]}  \\
\end{array}
\right),$$ and let us study two cases:\\\\
a). $\Lambda_u>0$ : In this case we have a lorentzian metric, and
$\omega_c$ takes its values in the lie algebra of
$\mathrm{SO(1,1)}\rtimes\mathrm{R}^2$.

This matrix valued one-form can be regarded as a
$\mathrm{SO(1,1)}\rtimes\mathrm{R}^2$ Cartan connection
~\cite{Sharpe} on the principal bundle
$\mathrm{SO(1,1)}\rightarrow \mathrm{P}\rightarrow M$ with
associated curvature $\Omega_c=d\omega_c+\omega_c\wedge\omega_c$
given by
\begin{equation}\Omega_c=\left(%
\begin{array}{ccc}
  0 & 0 & 0 \\
  T^1 & \Omega^1_1 & \Omega^1_2 \\
  T^2 & \Omega^2_1 & \Omega^2_2 \\
\end{array}%
\right)=\left(
\begin{array}{ccc}
  0 & 0 & 0 \\
  0 & -R & 0  \\
  0 & 0 & R \\
\end{array}
\right)\\\label{curv}\end{equation} where
$\Omega^i_j=d\omega^i_j+\omega^i_k\wedge\omega^k_j$ is the
standard curvature, and
\begin{equation}
R=-\frac{1}{a}\,a_{uu}\theta^1\wedge\theta^2
.\end{equation}\\
b). $\Lambda_u<0$ : In this case we have a riemannian metric, and
$\omega_c$ takes its values in the lie algebra of
$\mathrm{SO(2)}\rtimes\mathrm{R}^2$.

 This construction gives a $\mathrm{SO(2)}\rtimes\mathrm{R}^2$ Cartan
connection on the principal bundle $\mathrm{SO(2)}\rightarrow
\mathrm{P}\rightarrow M$ with associated curvature
$\Omega_c=d\omega_c+\omega_c\wedge\omega_c$ given by a similar
formula to (\ref{curv}).
\section{Conclusions}
In this note we show that, as it happens in NSF, all
two-dimensional riemannian and lorenztian metrics can be obtained
from the geometrical condition of a torsion-free connection.
Furthermore, we construct all associated Cartan connections to
these equations. This approach can be extended to the study of
third order ODE and a pair of 2nd order PDEs, but the choice of
the differential quadratic form $h$, is not \textit{a priori} so
clear. In fact, all these problems should be studied with the
Cartan's equivalence method~\cite{Olver} applied to differential
equation under a subgroup of contact transformations: the
canonical transformations. With this algorithmic method, the group
arising from these equations can be naturally obtained as the
group of allowed transformations in the study of the equivalence
problem. Work in this area has begun.
\section{Acknowledgments}
The author thanks Carlos Kozameh for the reading and revision of
this paper. The author is supported by CONICET.

\end{document}